# High-pressure synthesis of MgB$_2$ with addition of Ti

T.A. Prikhna [a], W. Gawalek [b], Ya.M. Savchuk [a], V.E. Moshchil [a], N.V. Sergienko [a], R. Hergt [b],

M. Wendt [b], S.N. Dub [a], Ch. Schmidt [b], J. Dellith [b], V.S. Melnikov [c], T. Habisreuther [b],

D. Litzkendorf [b], P.A. Nagorny [a]

[a] *Institute for Superhard Materials, 2, Avtozavodskaya Str., Kiev, 04074, Ukraine*

[b] *Institut für Physikalische Hochtechnologie, Winzerlaer Strasse 10, Jena, D-07745, Germany*

[c] *Institute of Geochemistry, Mineralogy and Ore-Formation, 34, Palladin Pr., Kiev, 02142, Ukraine*

**Abstract**

The MgB$_2$ – based material high-pressure synthesized at 2 GPa and 800$^{o}$C for 1 h from Mg and B (taken in the stoichiometry ratio of MgB$_2$) with addition of 2-10 wt% of Ti demonstrated the critical current density (j$_c$) higher than 100 kA/cm$^2$ at 33 K in 0 T field and at 20 K up to 3 T. At 20 K the critical current density higher than 10 kA/cm$^2$ was observed up to 5 T field. In the magnetic field up to 2 T high-pressure synthesized MgB$_2$ (with 10 % of Ti) at 20 K behaves in the same manner as Nb$_3$Sn at 4.2 K.

In XRD patterns of magnesium diboride with the Ti added, we observed no evidence of titanium diboride or unreacted titanium and only one compound with titanium was identified, namely, titanium dihydride TiH$_2$ (or more strictly TiH$_{1.924}$). The sample with the highest critical current density and irreversible field in the temperature range of 25-10 K contained some amount of pure Mg that was rather homogeneously dispersed in the material.

The critical currents and irreversible fields of magnesium diboride synthesized at high pressure with Ti added are higher than those in the case where Ta has been added.

Corresponding author: Dr.Sci.(Eng.) T.A. Prikhna, Institute for Superhard Materials,

2, Avtozavodskaya Str., Kiev, 04074, Ukraine, Fax:+380-44-468-86-25, E-mail address:

prikhna@iptelecom.net.ua

**Introduction**

We continue to study the influence of different additions on superconducting properties associated with pinning of high-pressure synthesized $MgB_2$. In our previous study it has been shown that the addition of Ta in amount of 2 or 10 wt% to the starting mixture of Mg and B (taken in the stoichiometric ratio of $MgB_2$) leads to an increase of critical current density and irreversible field of high-pressure synthesized $MgB_2$-based material Ref. [1]. We observed further improvement of the superconductive properties when Ti (10 wt%) was added to the starting Mg-B mixture instead of Ta. Here we discuss the results of studies of structural and superconductive characteristics of high-pressure synthesized $MgB_2$ samples with Ti additions (2 and 10 wt.%).

The structure of superconductive $MgB_2$ material turns out to be a rather complicated nanostructure. More and more investigators, Refs. [2-5], support the idea that the presence of oxygen in the structure of $MgB_2$ – based material can positively influence its critical current density. It has been shown in Refs [2, 3] that a thin film of $MgB_2$ alloyed with oxygen of the boron layer (with a c-axis parameter of 0.3547 nm larger than that (0.3521 nm) for bulk) exhibited the lower $T_c$ but had a higher slope $dH_{c2}/dT$ both in parallel and perpendicular field than the film with usual parameters. Also, the authors supposed that additional corpinning by the non-superconducting MgO particles can contribute to the total pinning force.

The results of an atomic resolution study, Ref.[4], have shown that the oxygen substitution occurs in the bulk of $MgB_2$ grains, to form coherently ordered $MgB_{2-x}O_x$ precipitates (≈20-100 nm in size) and that they act as pinning centers, thus increasing the critical current both at low and at high magnetization. The investigations of $MgB_2$ samples prepared by solid-state reaction, Ref.[5], using the high-resolution transmission electron microscopy, X-ray energy-dispersive spectroscopy, electron energy-loss spectroscopy and energy filtered imaging have shown that the $MgB_2$ matrix contains large amounts of coherent precipitates with a size ranging from about 5 nm up to 100 nm. The densities of the precipitates with a size ranging from 10 nm to 50 nm and from 50 nm to 100 nm were approximately $10^{15}/cm^3$ and $10^{14}/cm^3$, respectively. Traces of oxygen were seen in the $MgB_2$ matrix while the oxygen content was higher in the small precipitates of $Mg(B,O)_2$ and only magnesium and oxygen signals were seen in the large precipitates of MgO, Ref. [5]. Smaller precipitates had the same lattice structure and similar lattice parameters and the same crystal orientation as the $MgB_2$ matrix. The estimation of the precipitate effect on vortex pinning allowed the conclusion that precipitates larger than 5 nm are relevant pinning centers in $MgB_2$, Ref.[5]. The oxygen content increased while the boron content decreased with increasing precipitate sizes. The authors of Ref. [5] suggested the following formation of precipitates: oxygen dissolved in $MgB_2$ at a high temperature was later forced out to form $Mg(B,O)_2$ precipitates due to its lower solubility at lower temperatures; the precipitates were of the same basic structure as the $MgB_2$ matrix but with composition modification; long-term exposure to oxygen at high temperatures resulted in the transformation of $Mg(B,O)_2$ precipitates into MgO with a little change in precipitate sizes.

The Ti addition to Mg and B leads to an increase of the critical current density and irreversible field, Refs. [6-8]. The authors of Ref. [7] have found that Ti does not occupy the atomic site in the $MgB_2$ crystal structure, but forms a thin $TiB_2$ layer in the grain boundaries of $MgB_2$. Besides, $MgB_2$ grains are greatly refined by Ti doping, forming a strongly coupled nanoparticle structure. They concluded that the enhanced bonding between grains and the increased area of grain boundaries by forming the $MgB_2$ nanoparticle structure may be responsible for the enhancement of $j_c$ in Ti-doped $MgB_2$ bulk superconductors. By chemical vapor deposition of Ti and B on a commercial carbon-coated silicon carbide substrate, a starting fiber with Ti well dispersed in the B was obtained in Ref. [8]; these fibers in a Mg vapor were transformed into a superconductor capable of carrying 10 kA/cm$^2$ at 25 K and 1T. This superconducting magnesium diboride contained the intragranular precipitates that had appropriate size (from 1 to 20 nm) and spacing (5 times larger than the size of precipitates) for the pinning of vortices (Ref. [8]).

In our case, the structure of high-pressure synthesized $MgB_2$ bulk samples with Ti addition was investigated using X-ray diffraction, SEM (with microprobe analysis) and polarizing microscopy. We found only one Ti-containing compound, $TiH_{1.924}$, in the sample structure. The particle size of $TiH_{1.924}$ (from 1 to 10 µm) and distances between them seemed to be rather large to suppose that these particles can be vortex pinning centers. Besides, the increase in superconducting characteristics because of the Ti addition in high pressure-synthesized samples was not so pronounced as in the case of samples prepared by solid-state reaction at ambient pressure presented in Refs.[ 6,7], while the absolute values of critical currents in the field higher than 1 T for high-pressure synthesized samples were essentially higher than that of synthesized under the ambient pressure.

At present we are able to synthesize at high pressure (2 GPa) $MgB_2$ samples 30 mm in diameter and 15 mm in height of almost theoretical density and free of cracks (the preparation of larger samples is in progress). Samples of such a size can be used for practical application, for example, at 20 K, in different superconductive devices: electromotors and generators, energy storage systems, Maglev transport, contact-free bearings, etc.

## 2. Experimental

In the experiments on synthesis, metallic Mg chips and amorphous B (of 95- 97 % purity, MaTecK) have been taken in the stoichiometric ratio of $MgB_2$. To study the influence of Ti, the Ti metallic powder (of 99 % purity, MaTecK) or powder of $TiH_2$ has been added to the stoichiometric mixture of Mg and B in amounts of 2 or 10 wt%. The X-ray study of the initial Mg, Ti, Ta and B has shown that the materials contained no impurity phases with hydrogen (method accuracy being about 3-5%). Then we have mixed and milled the components in a high-speed activator with steel balls for 1-3 min. The obtained powder has been compacted into tablets.

High-pressure has been created inside high-pressure apparatuses (HPA) of the recessed-anvil and cube (six punches) types described elsewhere, Ref. [9]. The sample was in contact with a precompacted powder of hexagonal BN. Samples were synthesized at 800-900 $^o$C under a pressure of 2 GPa for 1 h.

The structure of the materials was studied using SEM and energy dispersive X-ray analysis, polarizing microscopy, and X-ray diffraction analysis. The $j_c$ was estimated from magnetization hysteresis loops obtained with an Oxford Instruments 3001 vibrating sample magnetometer (VSM) using Bean's model, Ref. [10].

3. **Results and discussion.**

Figures 1 and 2 show the critical currents vs. magnetic field at different temperatures estimated for high-pressure synthesized $MgB_2$ samples with Ti addition and for $MgB_2$ synthesized under ambient pressure with and without Ti addition. In Fig. 1 we compare our results obtained using a vibrating sample magnetometer (VSM) with the best data for bulk $MgB_2$ of several research groups given in the overview Ref. [11]. It should be mentioned that in magnetic fields up to 2 T, the high-pressure synthesized $MgB_2$-Ti material behaves at 20 K as $Nb_3Sn$ at 4.2 K and at 10 K its critical current density is higher in the fields up to 4 T than that of $Nb_3Sn$ at 4.2 K (Fig. 1). In Fig. 2 we compare the same our data with the data for the Ti-doped $MgB_2$ samples prepared by a solid-state reaction under the ambient pressure, Ref. [6]. High-pressure synthesized $MgB_2$-Ti demonstrated at 20 K in the fields 1-5 T the same characteristics as the material synthesized under the ambient pressure at 10 K (in magnetic fields above 5 T, the critical current density of the high-pressure synthesized material is even higher). All samples synthesized under high pressure show some decrease of $T_c$ (to 39-37 K).

In Fig. 3 the effects of Ti and Ta addition on superconductive properties of high-pressure synthesized samples are summarized. For each temperature and each magnetic field the best data of $j_c$ for the high-pressure synthesized material without and with Ti (Fig. 3a) or Ta (Fig.3b) addition are given. The effect of Ti-doping on superconductive properties of the high-pressure synthesized material appeared to be not so well pronounced as in the case of the material synthesized under ambient pressure when the increase of critical current density by two orders of magnitude has been observed, Ref. [6]. Somewhat lesser improvement of superconductive characteristics was observed when Ta was added to the high-pressure synthesized material (e.g., at 20 K the $j_c$ is higher than 100 $kA/cm^2$ up to 2.4 T with Ta and up to 3 T with Ti addition). It should be mentioned that in Fig. 3, we want to show "pure" influence of additions and because of this the amount of addition or the synthesis temperature were not taken into consideration, i.e. we want to show what characteristic we can get using high pressure only and high pressure together with Ti or Ta addition.

Figure 4 presents the results of the VSM and X-ray studies of samples high-pressure synthesized at different temperatures with additions of Ti and $TiH_2$. The amount of MgO has been approximately the same for all samples independently of the synthesis conditions and the amount of addition. Similar feature was observed in the case when Ta was added. As X-ray patterns show, there is no free Ti in the samples synthesized under 2 GPa for 1 h at 800-900 °C with 2 or 10 % of Ti and $TiH_2$ (or more strictly $TiH_{1.924}$) is the only one Ti-containing compound present in all samples.

The sample shown in Fig. 4a, that demonstrated the highest level of $j_c$ at 10-25 K in the external magnetic field higher than 1 T, contained some amount of free Mg. Mg seemed to be distributed homogeneously over the sample structure (according to the SEM study and microprobe analysis). The same picture was observed in the case of Ta addition: samples with the highest critical currents at 10-25 K contained free Mg in their structure.

The best superconducting properties in 10-25 K and 30-35 K temperature ranges were observed for the samples with 10 wt % addition of Ti synthesized for 1h under 2 GPa at $T_s$=800 °C (Fig. 4a) and $T_s$=900 °C (Fig. 4c), respectively. In the case when Ts=800 °C and the amount of Ti was 2 wt%, we observed the presence of $MgH_2$ in the sample structure (Fig.4b). When Ta was added, the higher amount of $MgH_2$ was found in the samples that have lower $j_c$ in the 10-30 K temperature range. When $T_s$=900 °C and the amount of Ti was 2 wt%, the lesser amount of $TiH_2$ (or more exactly, $TiH_{1.924}$) was observed in the material structure (Fig. 4d) than when 10 wt% Ti was added (Fig. 4c) and the jc of the samples with 2 wt% of Ti were lower as well. In the case of Ta, the higher amount of $Ta_2H$ was a characteristic feature of samples with higher $j_c$. The source of hydrogen can be atmosphere or moisture.

The $MgB_2$ – based material high-pressure synthesized at 2 GPa and 800°C for 1 h from Mg and B (taken in the stoichiometry ratio of $MgB_2$) with addition of 2-10 wt% of Ti demonstrated the critical current density ($j_c$) higher than 100 kA/cm$^2$ at 33 K in 0 T field, at 30 K up to 0.8 T field, at 25 K up to 1.7 T, at 20 K up to 3 T and at 10 K up to 4.5 T. At 20 K the critical current density higher than 10 kA/cm$^2$ was observed up to 5 T field. At 35 K the critical current density of the material reached 50 kA/cm$^2$ and the irreversible field at this temperature was 1 T.

When we added TiH$_2$ instead of Ti to the Mg and B mixture, the superconductive characteristics were drastically depressed (Fig. 4e). Fig. 4e shows the sample with TiH$_2$ added that had the highest superconductive characteristics among all the samples being studied (with 2 and 10 wt% addition of TiH$_2$, synthesized at 800, 900 and 950 $^o$C under 2 GPa for 1h). The X-ray pattern (Fig.4e) shows that it contains a great quantity of TiH$_2$ (or TiH$_{1.924}$).

Figs. 5 and 6 show the SEM patterns of the structures of high-pressure synthesized MgB$_2$ (pure and with additions of Ti and Ta). Structures shown in Fig. 6 were photographed at the same magnification. Of course, we unable to detect hydrogen or hydrides using SEM and microprobe analysis and we marked grains of hydrides in the photos based on XRD analysis and microprobe detection of Ta or Ti.

As is seen from Fig. 5, the grains contained Ti seem too large to be pinning centers in MgB$_2$. The SEM study using microprobe analysis has shown that there were no diffusion of Ti or Ta into the "matrices" of the samples. The lattice parameters of MgB$_2$ remained unchanged after high-pressure synthesis with Ti or Ta additions. Usually, the so-called "matrix" of a sample contains Mg, O and B. The "matrix" has very fine and complicated structure so it has not been possible to examine it in details by SEM or, as it were shown by the authors of Refs. [2,4,5,7,8], even by TEM of high resolution.

A comparison between the structures with 10 wt% of Ti and TiH$_2$ additions shown in Fig. 6a and Fig. 6c, respectively, gives grounds to conclude that the structure shown in Fig. 6c is less dense than that in Fig. 6a. The most likely explanation can be that hydrogen that contained in TiH$_2$ or trapped in the precursor material during its preparation (milling, compacting, etc) is liberated under high pressure-high temperature conditions, thus preventing the densification of the structure. The low density of the structure can be one of the reasons why the critical current density decreases. We understand the drawbacks and contradictions of above explanation. Because when there were no additions to Mg and B, the synthesized samples were nearly theoretically dense and some amount of MgH$_2$ was detected in their structure, Ref. [1]; when we added Ti or Ta to the Mg and B mixture the hydrides of Ti or Ta were formed, but, when we added TiH$_2$, under the same conditions hydrogen seems to be liberated from TiH$_2$, provoking a reduction of the material density, and in the produced material (as X-ray shown, Fig. 4e) there were no free Ti and MgH$_2$ but only TiH$_2$ (TiH$_{1.924}$), MgB$_2$ and MgO were detected. In this connection we hope that our future experiments will give us enough information to understand the observed phenomena.

Unfortunately, up to now we cannot suggest the legible mechanism of Ti and Ta positive influence on the superconductive properties of high-pressure synthesized MgB$_2$. We have observed a lot of evidences of the negative effect on superconductivity of hydrogen present in the form of MgH$_2$ or of free hydrogen liberated during the process. Besides, in our first experiments, in which we used initial boron with a great amount of H$_3$BO$_3$ impurity, the critical currents were lower by an order of magnitude than in case, when we used pure boron (but even in the case of boron contained the H$_3$BO$_3$ impurity, the positive effect of Ta was already observed). Most likely, the positive influence of Ti and Ta are due to the absorption of hydrogen. When we add Ta, Ta$_2$H and sometimes TaH form and in the case of Ti, TiH$_{1.924}$ forms. So, the higher effect of Ti as compared to Ta can be explained by the higher absorptivity of Ti with respect to hydrogen.

The main difference between the structures shown in Figs. 6b, d, e, f, g is the amount or the density of distribution and the size of black Mg-B (most likely MgB$_2$) inclusions. The comparison of structures and sample properties support our suggestion made in Refs. [1, 12] that the critical current density of MgB$_2$ is strongly influenced by the density of distribution of black Mg-B inclusions. The basic effect on the increase of the amount of Mg-B inclusions in high-pressure synthesized samples is exerted by the reduction of the synthesis temperature. It seams that additions of Ti and Ta influence the density of Mg-B inclusions as well.

## 4. Conclusion.

The data given in the present paper show the positive effect of Ti addition on superconductive characteristics of high-pressure synthesized $MgB_2$. The multiphase and very fine structure of samples gives no way of proposing the legible mechanism of this influence. But there are a lot of evidences that the basic effect of Ti can be due to the absorption of hydrogen impurity and possibly due to the fact that Ti promotes the formation of small (nanosized) Mg-B ($MgB_2$) inclusions thus causing the increase in critical current density. The more pronounced Ti effect as compared with that of Ta may be explained by the higher absorptivity of Ti. The high-pressure synthesized $MgB_2$ samples with Ti addition were approximately theoretically dense and exhibited the critical current densities in the magnetic fields above 1 T higher than those that were reported in literature for bulk $MgB_2$.

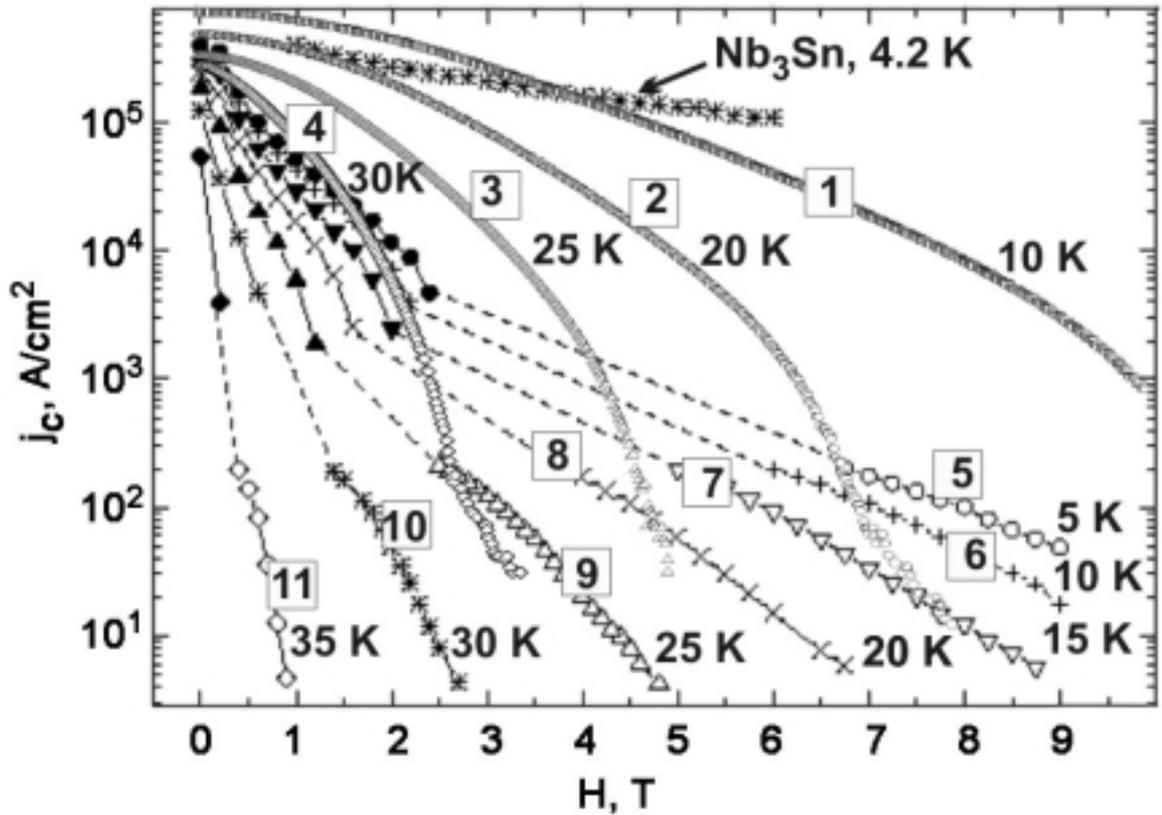

Figure 1. Critical current density vs. magnetic field obtained using vibrating sample magnetometer and the Bean model of high-pressure synthesized $MgB_2$-Ti samples at 10, 20, 25 and 30 K (curves 1-4, respectively). The data for $Nb_3Sn$ at 4.2 K and for polycrystalline $MgB_2$ and wire segments synthesized from Mg and B under the ambient pressure at 5, 10, 15, 20, 25, 30, 35 K (curves 5-11, respectively), taken from overview Ref. [11], are given in the same Figure for comparison. (The upper parts of curves 5-10, i.e. higher values of $j_c(H,T)$, were inferred from magnetization data using the Bean model and the lower parts (lower values of $j_c$) were directly measured via V(I) measurements).

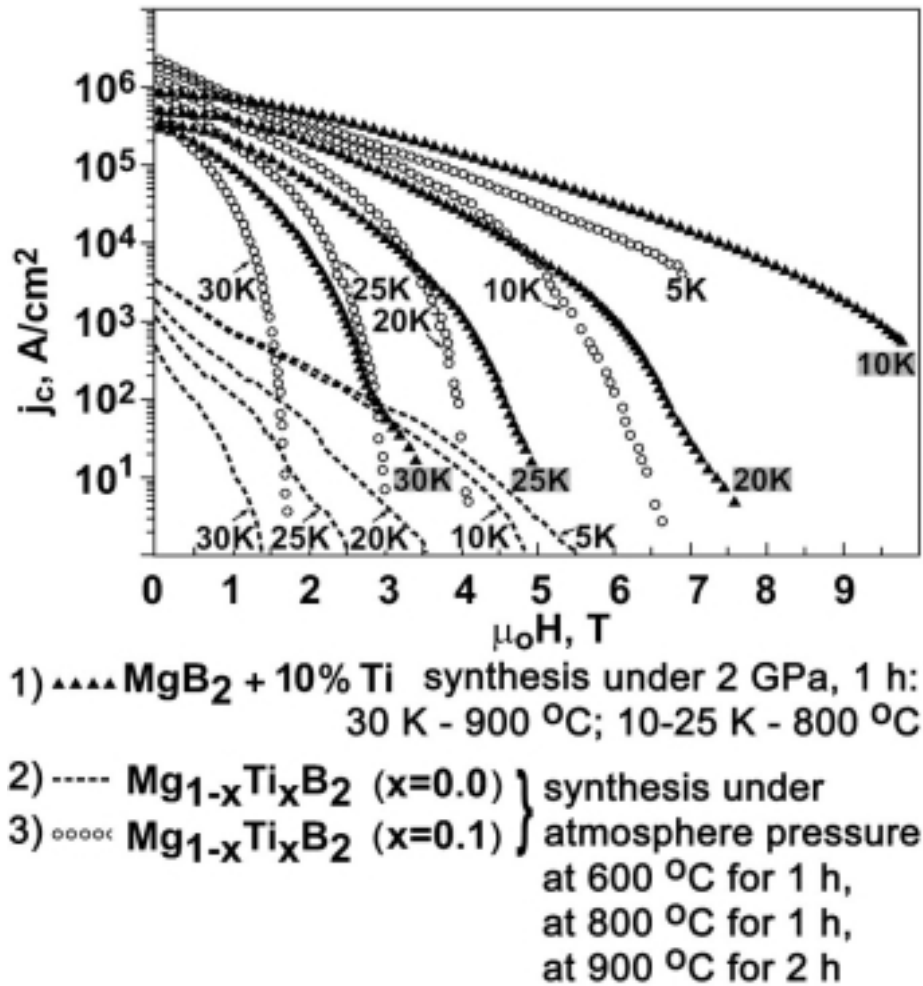

Figure 2. Critical current density vs. magnetic field (obtained from magnetization data) of high-pressure synthesized $MgB_2$-10%Ti samples at 10, 20, 25 and 30 K (curves marked by solid triangulates (1) and of $MgB_2$ samples with Ti doping ($Mg_{1-x}Ti_xB_2$) for x=0 (curves marked by dashed lines (2)) and x=0.1 (marked by empty circularizes (3)) taken from Ref.[6].

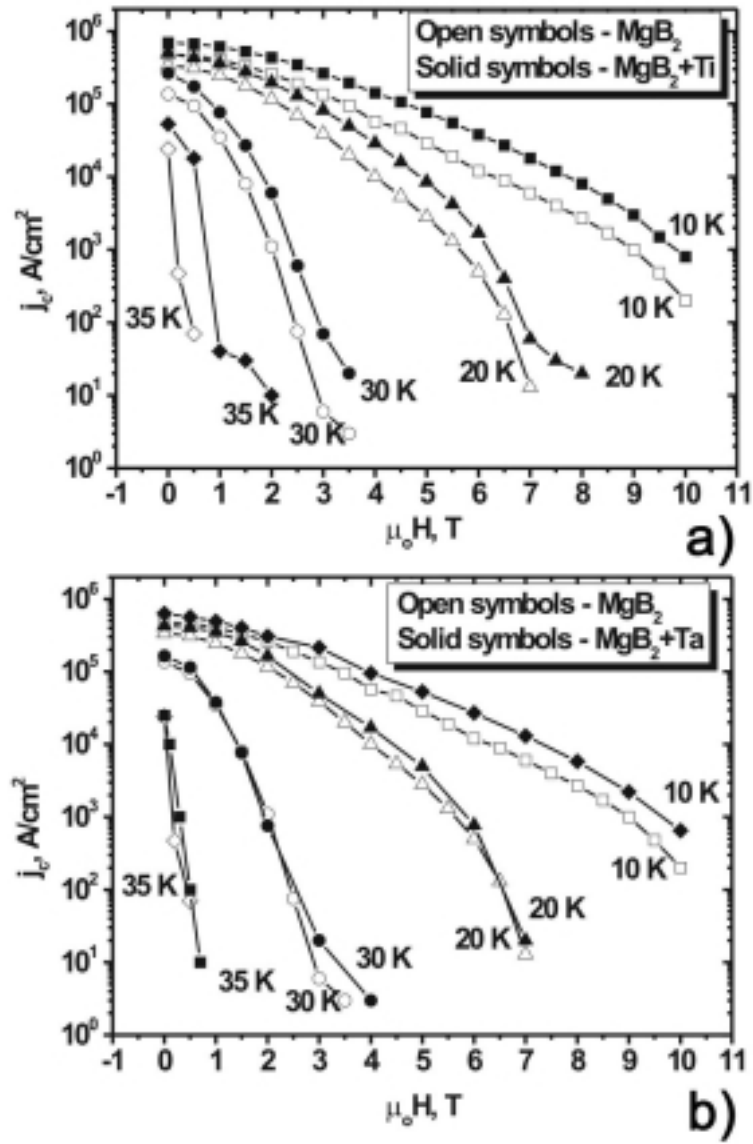

Figure 3. Comparison of the critical current density, $j_c$, vs. magnetic field, $\mu_0 H$, for the high-pressure synthesized $MgB_2$ samples with that for high-pressure synthesized $MgB_2$-Ti (Fig. 3a) and $MgB_2$-Ta (Fig 3b) samples. For each temperature and each magnetic field, the best data of $j_c$ for high-pressure synthesized material without and with Ti or Ta addition are given.

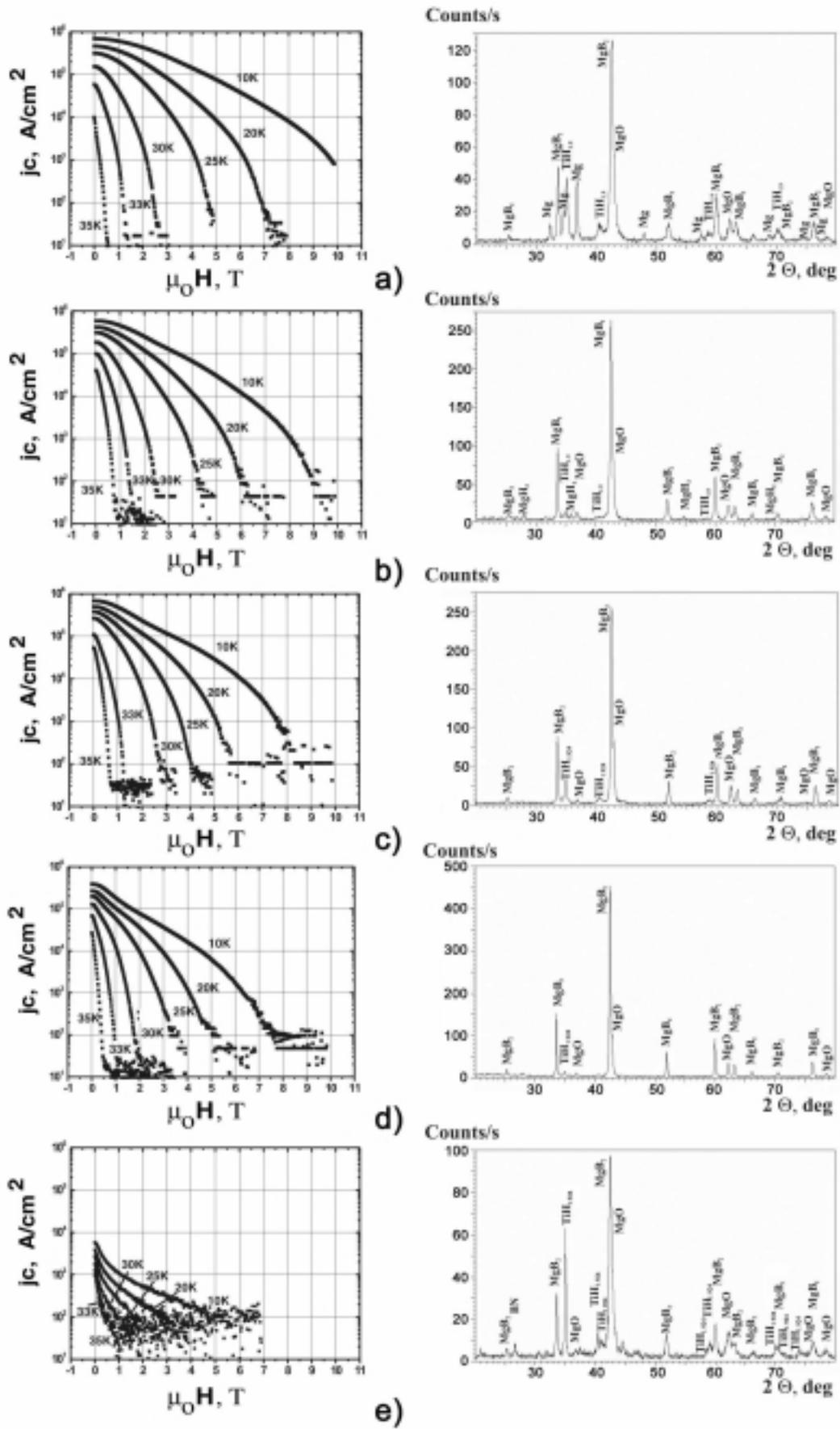

Figure 4. VSM and X-ray study of the high-pressure synthesized MgB$_2$ samples in contact with BN: with Ti additions (a-d): (a) 10 wt % Ti, 2 GPa, 800 $^o$C, 1h, ; (b) 2 wt% of Ti, 2 GPa, 800 $^o$C, 1h ; (c) 10 wt% Ti, 2 GPa, 900 $^o$C, 1h ; (d) 2 wt% of Ti, 2 GPa, 900 $^o$C, 1h; and with TiH$_2$ addition (e) 10 wt% TiH$_2$, 2 GPa, 950 $^o$C, 1h.

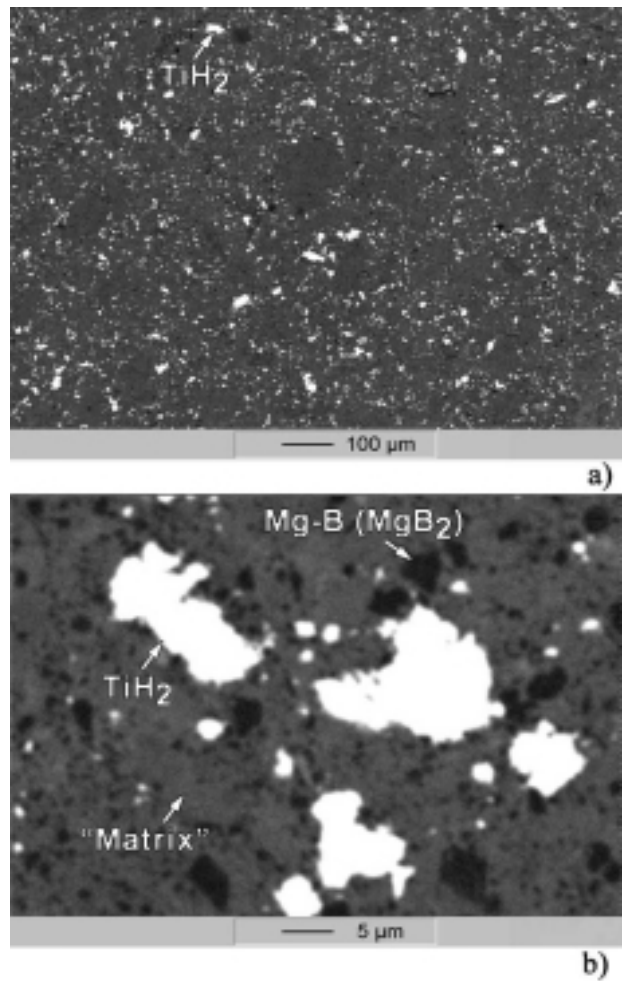

Figure 5. Structure of high-pressure synthesized MgB$_2$ with 10 wt. % of Ti at 2 GPa, 800 $^o$C for 1 h at different magnifications (obtained using SEM, "composition" – back scattering electron images, in which the heavier element, the brighter it looks).

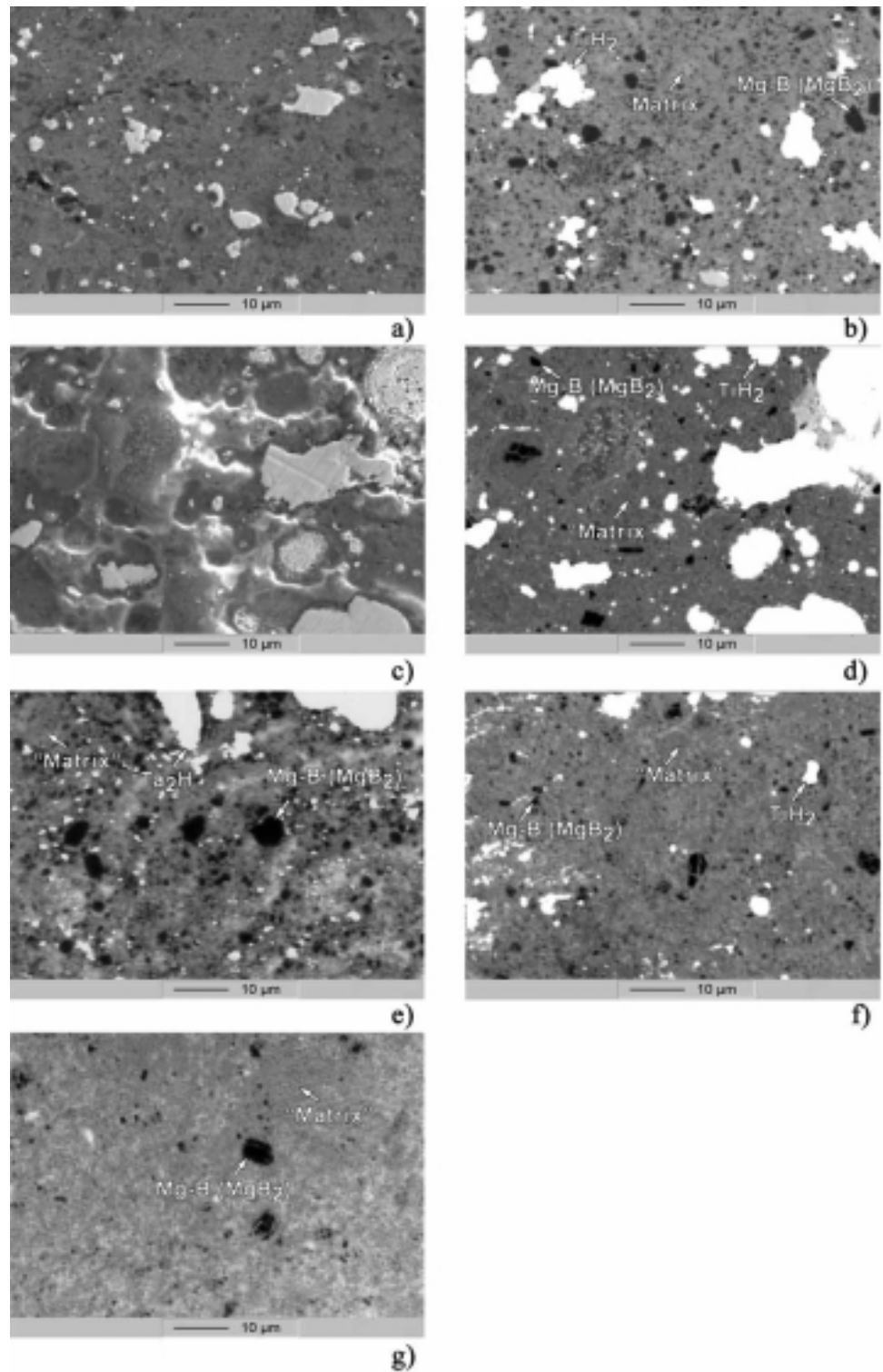

Figure 6. Structure of high-pressure synthesized MgB$_2$ with Ti, Ta and TiH$_2$ addition obtained by SEM at the same magnification: (a,b) 10 wt% Ti, 2 GPa, 800 °C, 1h: SEI (a) and COMPO (b) images (not the same area); (c, d) 10 wt % TiH$_2$, 2 GPa, 950 °C, 1 h: SEI (c), COMPO (d) images (the same area); 10 wt% Ta, 2 GPa, 800 °C, 1h, COMPO image; 2 wt% Ti, 2 GPa, 800 °C, 1h, COMPO image; MgB$_2$ without additions, 2 GPa, 800 °C, 1h, COMPO image.